# Novel magnetic behavior of single crystalline $Er_2PdSi_3$

K.K. Iyer, P.L. Paulose, E.V. Sampathkumaran
*Tata Institute of Fundamental Research, Homi Bhabha Road, Colaba, Mumbai – 400005, India*

M. Frontzek, A. Kreyssig, M. Doerr, M. Loewenhaupt
*TU Dresden, Institut für Festkörperphysik, D-01062 Dresden, Germany.*

I. Mazilu, G. Behr, and W. Löser
*Leibniz-Institut für Festkörper- und Werkstoffforschung Dresden, Postfach 270116, D-01171 Dresden, Germany*

## Abstract

We report the results of ac and dc magnetic susceptibility ($\chi$) and electrical resistivity ($\rho$) measurements on the single crystals of $Er_2PdSi_3$, crystallizing in an $AlB_2$-derived hexagonal structure, for two orientations H//[0001] and H//[$2\bar{1}\bar{1}0$]. **For H//[0001],** there are apparently two magnetic transitions as revealed by the $\chi_{ac}$ data, one close to 7 K attributable to antiferromagnetic ordering and the other around 2 K. However, **for H // [$2\bar{1}\bar{1}0$],** we observe additional features above 7 K (near 11 and 23 K) in the plot of low-field $\chi$(T); also, there is no corresponding anomaly in the $\rho$(T) plot. In this respect, the magnetic behavior of this compound is novel, particularly while compared with other members of this series. The features in $\chi_{ac}$ respond differently to the application of a small dc magnetic field for the two directions. As far as low temperature (T= 1.8 and 5 K) isothermal magnetization (M) behaviour is concerned, it exhibits meta-magnetic-like features around 2 kOe saturating at high fields for the former orientation, whereas for the latter, there is no saturation even at 120 kOe. The sign of paramagnetic Curie temperature is different for these two directions. Thus, there is a strong anisotropy in the magnetic behavior. However, interestingly, the $\rho$(T) plots are found to be essentially isotropic, with the data revealing possible formation of magnetic superzone formation below 7 K.

Corresponding author. E.V. Sampathkumaran: E-mail address: sampath@tifr.res.in

Considering interesting properties exhibited (see, for instance, Refs. 1-3 and references cited therein) by the polycrystalline samples of ternary intermetallic compounds derived from $AlB_2$-type hexagonal structure, we have been carrying out systematic magnetic and transport measurements on the single crystalline forms of the compounds of the series $R_2PdSi_3$ (Ref. 4). The studies revealed further novel behavior for the single crystalline form [5-10]. In particular, we have noticed a small degree of anisotropy in some properties for the Ho and Dy cases [8, 10], while for some others [5,7,9,11] (e.g., R= Tb), the observed anisotropy is stronger, surprisingly even for R= Gd, which is a S-state ion. In addition, probably in all these compounds, there appears to be another magnetic transition [1-3, 5-10] at a temperature below the initial magnetic ordering temperature; this second transition, though exhibits "anisotropic spin-glass-like" behaviour in macroscopic magnetic measurements, for instance, in the Tb case [9], does not appear to arise from a classical spin-glass behavior as deduced from recent neutron diffraction data [11]. We have extended single crystal investigations to another heavy rare-earth member of this series, viz., $Er_2PdSi_3$, and some of the features reported in this article present an interesting situation. It may be mentioned that previous investigations on the polycrystalline samples reveal [4,12] that this compound undergoes antiferromagnetic ordering with a sinusoidally modulated magnetic structure below 7 K with the magnetic moment lying parallel to c-axis.

Single crystals of $Er_2PdSi_3$ with 6 to 8 mm in diameter and 25 to 50 mm in length were grown by a vertical floating zone technique with optical heating. The growth process proceeded in a vacuum chamber under flowing Ar atmosphere purified with a Ti-getter furnace. Further details of crystal growth and characterization by Laue method will be discussed elsewhere [13]. The pieces with requisite dimensions for measurements were cut for two orientations $[2\bar{1}\bar{1}0]$ and [0001] of the rods. The magnetization (M) measurements (1.8 - 300 K) were performed employing a commercial superconducting quantum interference device (SQUID, Quantum Design) as well as a vibrating sample magnetometer (Oxford Instruments). The same SQUID magnetometer was employed to take ac magnetic susceptibility ($\chi_{ac}$) data below 25 K at various frequencies. The electrical resistivity (ρ) measurements were performed by a conventional four-probe method employing spring-loaded sharp gold-plated copper tips for making electrical contacts with the sample.

The temperature (T) dependence of dc magnetic susceptibility ($\chi_{dc}$) at low temperatures taken in the presence of magnetic fields (H) of 5 kOe and 100 Oe for the two orientations are shown in figures 1 and 2 respectively. For both the orientations, the plots of inverse $\chi_{dc}$ versus T taken with H= 5 kOe is linear above 70 K. The value of the effective moment (9.6 $\mu_B$/Er) obtained from the high temperature linear region is very close to that expected for trivalent Er ions. The $\chi_{dc}$ shows a continuous (but weak) deviation from the Curie-Weiss law as the T is lowered below 70 K for the H//$[2\bar{1}\bar{1}0]$ orientation. For H//[0001], the Curie-Weiss law however extends down to 30 K. We attribute the deviation from the linear behavior to crystal-field and/or short range order effects. In the following, we bring out the anisotropy in the $\chi_{dc}$ behavior including the anisotropic onset of short range magnetic correlations:

For **H//[0001]**, the Curie-Weiss temperature ($\theta_p^c$) obtained from the linear region above 100 K is found to be 16 K, as though the magnetic coupling along c-direction is ferromagnetic. In the $\chi$(T) plot recorded at H= 100 Oe (Fig. 2), there is a distinct peak at 7 K; however, the data recorded in 5 kOe (see Fig. 1 inset) show the peak at a lower temperature (close to 6 K) which establishes that the magnetic ordering is of an antiferromagnetic-type. We have also tracked the $\chi_{dc}$ behavior for the field-cooled (FC) state of the sample for H= 100 Oe (see Fig. 2) and we find that there is essentially no bifurcation of zero-field-cooled (ZFC) and FC curves which is a sufficient proof for the absence of spin-glass freezing at 7K.

For **H//[2$\bar{1}\bar{1}$0]**, the dc $\chi$ behavior is found to be completely different. The magnitude of $\chi$ is much smaller at low temperatures, for instance, about an order of magnitude less for the data taken with H= 5 kOe. The sign of $\theta_p^a$ (-8 K) above 100 K is negative, in contrast to the positive sign of $\theta_p^c$ observed for H//[0001]. The asymptotic Curie temperature for the system ($\theta_p^c + 2\theta_p^a$) is near zero pointing towards a coexistence of antiferro- and ferromagnetic interactions. The sign change of the asymptotic Curie temperature arises from the crystal-field anisotropy. In the data taken with H= 100 Oe (see figure 2), though two magnetic transitions are visible below 7 K as for H//[0001], there are multiple peaks, shoulders, or dips, e.g., around 11, 7, and 5 K; in addition, there is a bifurcation of zero-field-cooled and field-cooled curves around 23 K. Such multiple features are new observations for single crystals of this series, while others show different effects in the paramagnetic state [5-11]. The appearance of this 23K-bifurcation in the absence of a similar feature for H//[0001] and of a corresponding anomaly in electrical transport data (see below) is puzzling. It is possible that this temperature represents the onset of two-dimensional magnetic fluctuations (incipient low-dimensional magnetic order?).

Similar differences for the two orientations are seen in the $\chi_{ac}$ data taken at several frequencies, the results of which are shown in figure 3. While the results confirm the features from the $\chi_{dc}$ data with respect to the existence of magnetic transitions, in the sense that, for $H_{ac}$//[0001], there are two magnetic transitions, while for $H_{ac}$//[2$\bar{1}\bar{1}$0] there are multiple peaks/shoulders (as marked by short vertical arrows in figure 3) in $\chi$'(T) plots in zero field. We do not find any frequency dependence of the peak temperature for the real part ($\chi$') for the peaks above 4 K for both the directions, establishing that these transitions are truly not of a spin-glass type. Absence of a peak in the imaginary part ($\chi$") for $H_{ac}$//[0001] is in addition an argument in favor of this conclusion. However, below 4 K, there is a noticeable frequency dependence with significant signal for $\chi$''. Thus, for instance, for $\chi$' of $H_{ac}$//[0001], the upturn for the lowest frequency of 1.2 Hz persists down to 1.8 K, whereas a distinct peak is clearly visible around 2.5 K for the highest frequency of 1222 Hz. This appears to be true for $\chi$' of $H_{ac}$//[2$\bar{1}\bar{1}$0] as well. Thus, this property is isotropic unlike in $Tb_2PdSi_3$ [Ref. 9], thereby implying that the magnetic characteristics for the second transition below the Néel temperature are different for these two compounds. We would like to mention that the isothermal dc magnetization at 1.8 K falls to negligibly small values after dc H is switched off as one would expect for an antiferromagnet. The characteristic decay behavior of a spin-glass system is not observed. We have also taken the $\chi_{ac}$ data in the presence of a small dc magnetic field (which induces the metamagnetic-like transition for H//[0001]), say at 3 kOe, in the same direction as that of the ac field. It is clear from the figure 3 that the presence of this dc field suppresses the feature around 1.8 K for $H_{ac}$//[0001], but not for H//[2$\bar{1}\bar{1}$0]. It is possible that this suppression is due to a shift of the 2K-transition to a higher temperature (thereby merging with the features at higher temperatures, discussed below) for H//[0001]. At higher temperatures, the two directions respond differently to the application of this dc field. Thus, for $H_{dc}$//[$2\bar{1}\bar{1}0$], the multiple peaks in χ'(T) vanish resulting in a broad feature peaking around 12 K without any frequency dependence; χ''(T) is featureless. However, for H//[0001], the feature for the higher temperature transition (in the vicinity of 7 K) persists, but the peak is broadened with a shift to a lower temperature (to 5 K) as in the case of 5kOe-$\chi_{dc}$ behavior (Fig. 1); it is also to be noted that χ'' component interestingly also shows up a signal at the same temperature range; there is also a significant frequency dependence of $\chi_{ac}$ for this direction in the presence of 3 kOe. Thus, it appears that the application of a dc field of the order of 3 kOe induces *anisotropic* ac χ features, a behavior different from that observed in other members of this series. It is of interest to focus future investigations to understand this aspect. Finally, the values of χ'' for 1222 Hz for the hard axis are enhanced while compared with the values at other frequencies, the origin of which is not clear to us at present.

We now discuss the results of isothermal magnetization behavior at 1.8, 5 and 15 K, shown in figures 4 and 5. It is transparent from figure 4 that the M(H) plots are qualitatively different for the two orientations, revealing strong anisotropy. There is a sharp rise for initial applications of H for H//[0001] at 1.8 and 5 K, while the rise remains sluggish up to 120 kOe for H//**[$2\bar{1}\bar{1}0$]**. For the former orientation, at high fields, there is a saturation of M (to a value close to that of free-ion), suggestive of ferromagnetic coupling in the presence of such high fields, whereas, for the latter, M does not saturate at all with the value at 120 kOe falling well below that for H//[0001] establishing the hard direction. A careful look at the data of figure 4 in the low-field range for H//[0001] at 1.8 and 5 K indicated a curvature around 2 kOe. Therefore, we have performed hysteresis measurements by focusing in the field-range below 5 kOe (see Fig. 5). It is clear from the figure 5 that there is a metamagnetic-like curvature beyond about 2 kOe at both 1.8 and 5 K, which actually reveals antiferromagnetic coupling in zero magnetic-field along c-axis, despite positive values of $\theta_p^c$. However, it is fascinating to note that, at intermediate fields (2 to 9 kOe), hysteretic behavior is observed at 1.8 K, but essentially negligible at 5 K (Fig. 5). This can be explained by proposing that the metamagnetic-like transition could be first-order-like (broadened by disorder) at 1.8 K, and that the magnetic structure possibly gets modified in a subtle way as the T is decreased from 5 K to very low temperatures, consistent with the conclusions from the $\chi_{ac}$ data, presented above. For **H//[$2\bar{1}\bar{1}0$]**, in the magnetically ordered state, there is a weak upward curvature beyond 40 kOe as though there is a tendency for spin-reorientation effects at higher fields (see the deviation from the dashed line for 1.8K-curve shown in figure 4, extrapolated from low fields). The observed features overall reveal considerable degree of anisotropy in magnetic behavior, possibly due to both the crystal-field and the magnetic interaction.

In Fig. 6, we show the T dependence of ρ, which interestingly reveals essentially isotropic electrical conductivity. In the paramagnetic state, the ρ for both the directions of excitation current (I) with respect to the two orientations of the sample exhibits metallic behavior (that is, positive dρ/dT). The order of magnitude of the values for both the directions is the same. The ρ-curve exhibits an upturn at the onset of magnetic transition, as though magnetic Brillouin-zone boundary gaps are formed below 7 K, consistent with the modulated magnetic structure proposed by neutron diffraction measurements [12]. It should be noted (see the inset of figure 6) that there is no worthwhile feature around 23 K - the temperature at which the low-field magnetization data show magnetic anomalies - for I//$[2\bar{1}\bar{1}0]$.

Summarizing, we report the following interesting and novel findings on the single crystal of $Er_2PdSi_3$: (i) This compound exhibits strongly anisotropic magnetic susceptibility and isothermal magnetization behavior, however exhibiting essentially isotropic behavior in the temperature dependent electrical resistivity. (ii) There are apparently two magnetic transitions (7 K and < 2K) for H//[0001], whereas, for H//$[2\bar{1}\bar{1}0]$ interestingly, multiple features are seen in the $\chi(T)$ plots with the onset of magnetic anomalies around 23 K. The origin of 23K-feature in the low-field $\chi(T)$ plots for H//$[2\bar{1}\bar{1}0]$ alone is surprising and we attribute it at present to the existence of two-dimensional short range correlation effects considering the layered nature of the crystal structure. It would be rewarding to perform careful neutron diffraction experiments as a function of temperature on the single crystals of this compound to understand this aspect better. (iii) Unlike the Tb compound [9], the frequency dependence of $\chi_{ac}$ around 2 K is isotropic in zero-field, whereas the application of a magnetic field of about 3 kOe apparently induces anisotropy below around 5 K. In these respects, this compound exhibits unusual properties among this class of compounds.

The work is partly supported by DFG within the program of the Sonderforschungsbereich 463 'Rare Earth-Transition-Metal-intermetallic compounds: Structure, Magnetism, Transport'.

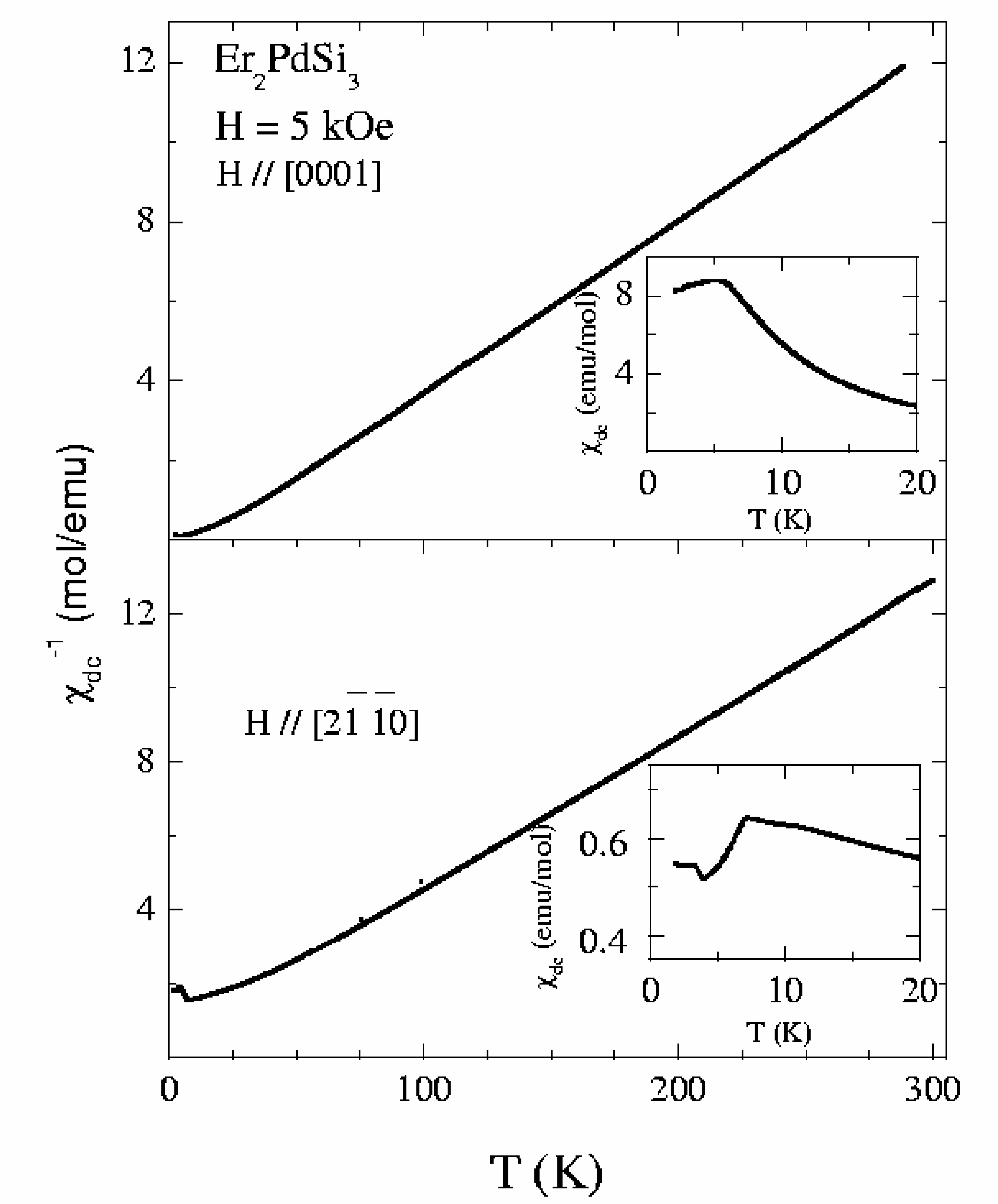

Figure 1. Inverse dc susceptibility as a function of temperature for two orientations of the single crystal of $Er_2PdSi_3$ with respect to the magnetic field of 5 kOe. The low temperature $\chi$ behavior is shown in the inset.

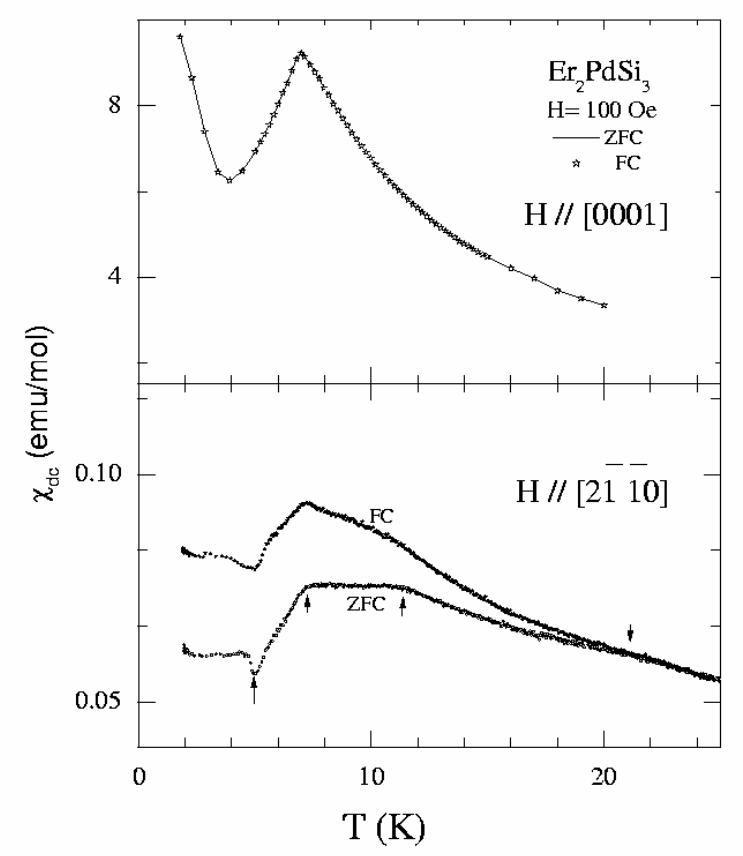

Figure 2. The low temperature magnetic susceptibility behavior of $Er_2PdSi_3$ for two orientations of the single crystal measured in a magnetic field of 100 Oe. ZFC and FC represent zero-field-cooled and field-cooled states of the specimens to 1.8 K and the curves corresponding to these two situations overlap for H//[0001]. Vertical arrows mark the temperatures at which there are clear shoulders/transitions, discussed in the text.

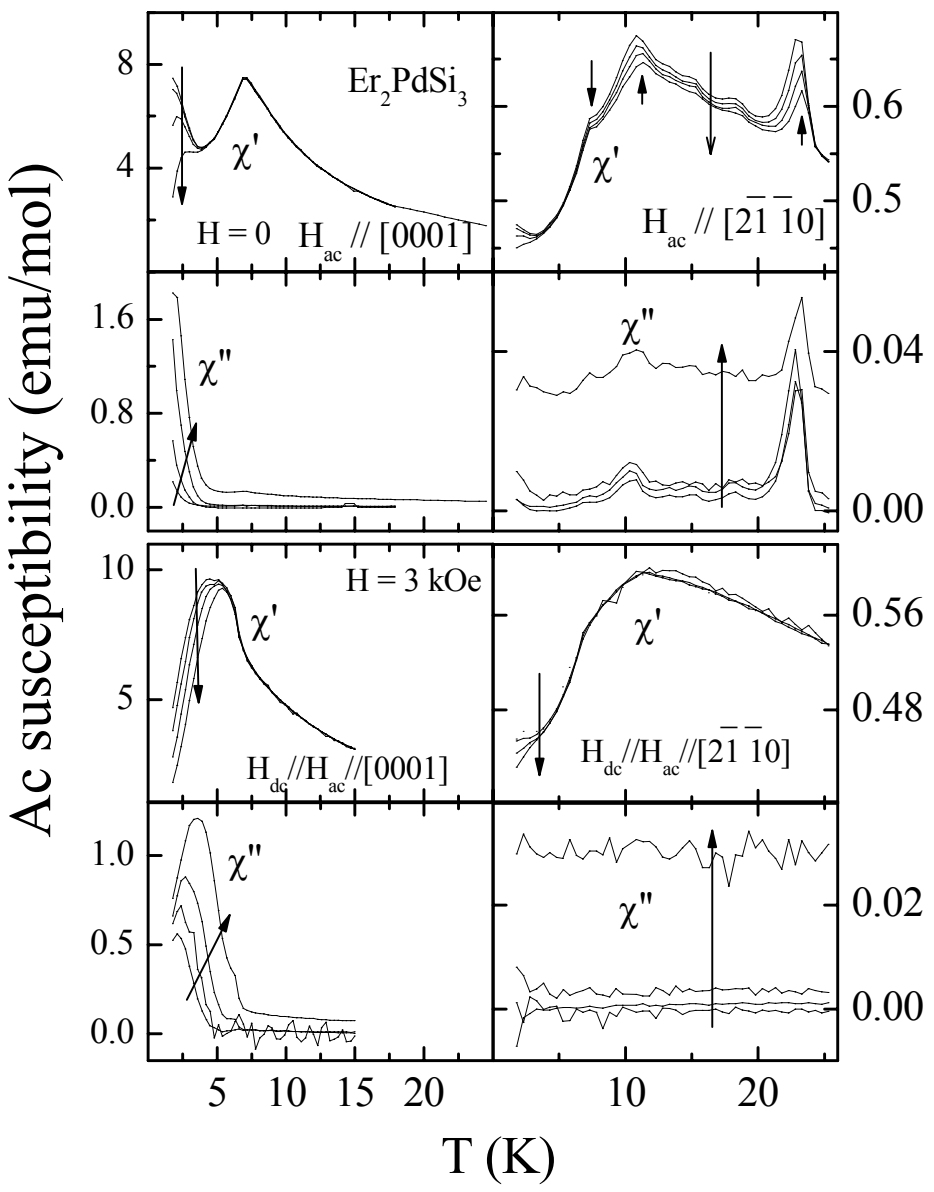

Figure 3. Real (χ') and imaginary (χ'') parts of ac magnetic susceptibility of $Er_2PdSi_3$ for two orientations with respect to the direction of the ac field (1 Oe), recorded at various frequencies. The top and bottom half correspond to zero and 3 kOe externally applied dc magnetic field for the directions specified in the figures. While small arrows in one of the figures mark the temperatures at which the features discussed in the text are seen, the long arrows are drawn to show the way the curves move with increasing frequency (1.2, 12, 122, 1222 Hz).

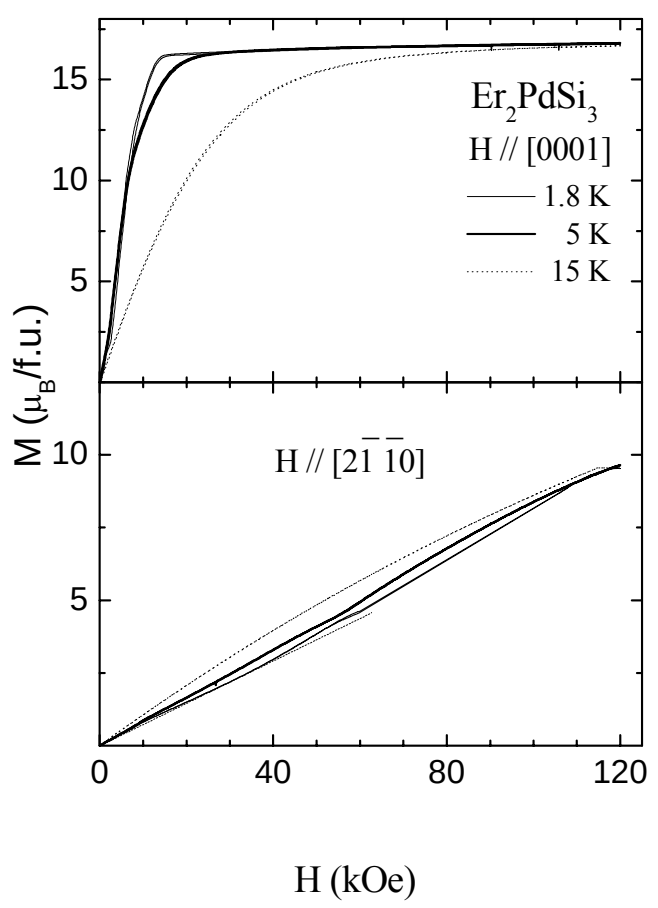

Figure 4. Isothermal magnetization per formula unit (f.u.) at 1.8, 5 and 15 K for two orientations of the single crystal of $Er_2PdSi_3$. The values for increasing and decreasing fields nearly overlap, except perhaps a hysteresis effect at intermediate fields for H//[0001] as described in the text. A dashed line for 1.8 K curve of H//[2$\bar{1}\bar{1}$0] is drawn by extrapolation from low-field linear region.

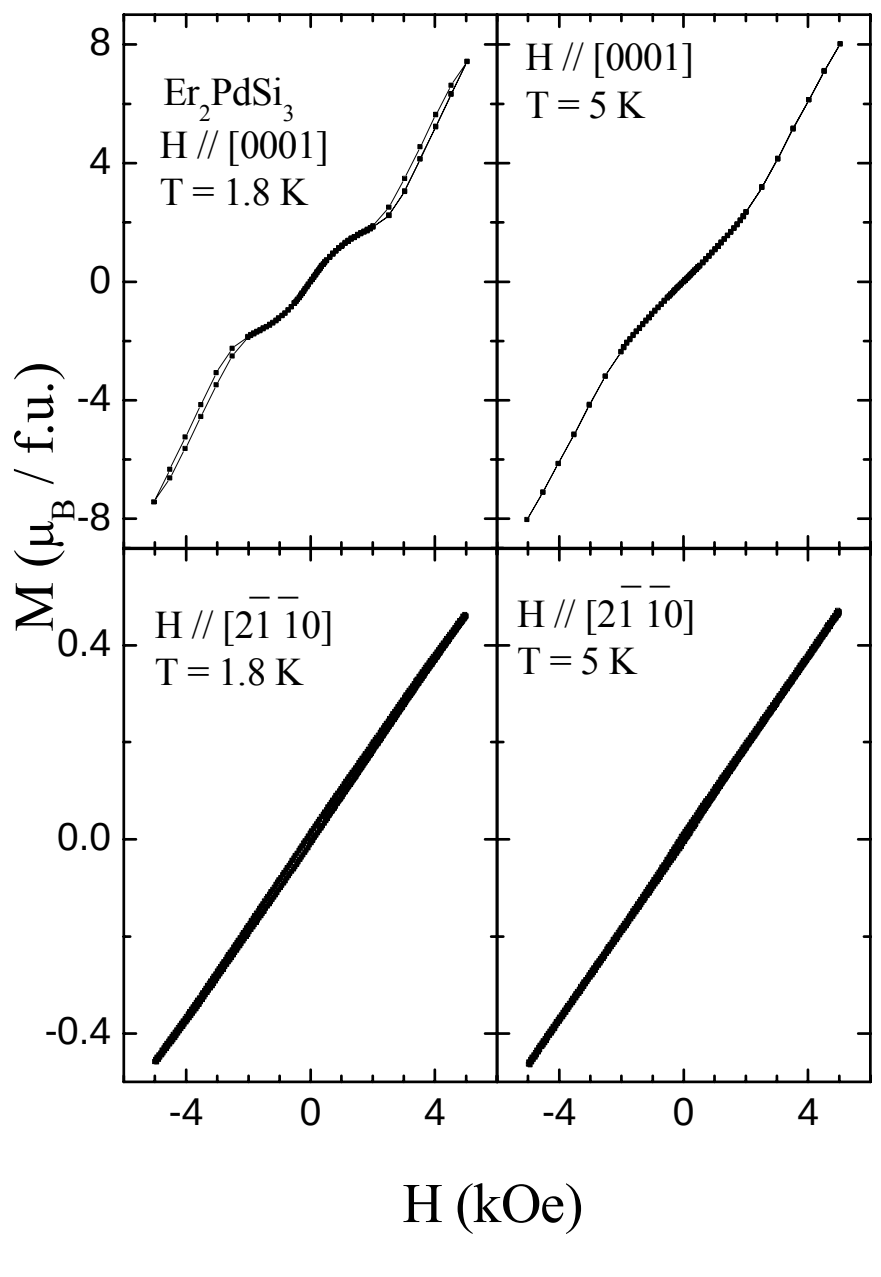

Figure 5. Low field magnetic hysteresis loops at 1.8 and 5 K to highlight the existence of metamagnetic transitions around 2 kOe for H//[0001] in $Er_2PdSi_3$. The lines drawn through the data points serve as guides to the eyes. The data points below 2 kOe in the top figures have been taken more densely to explore low-field features more closely.

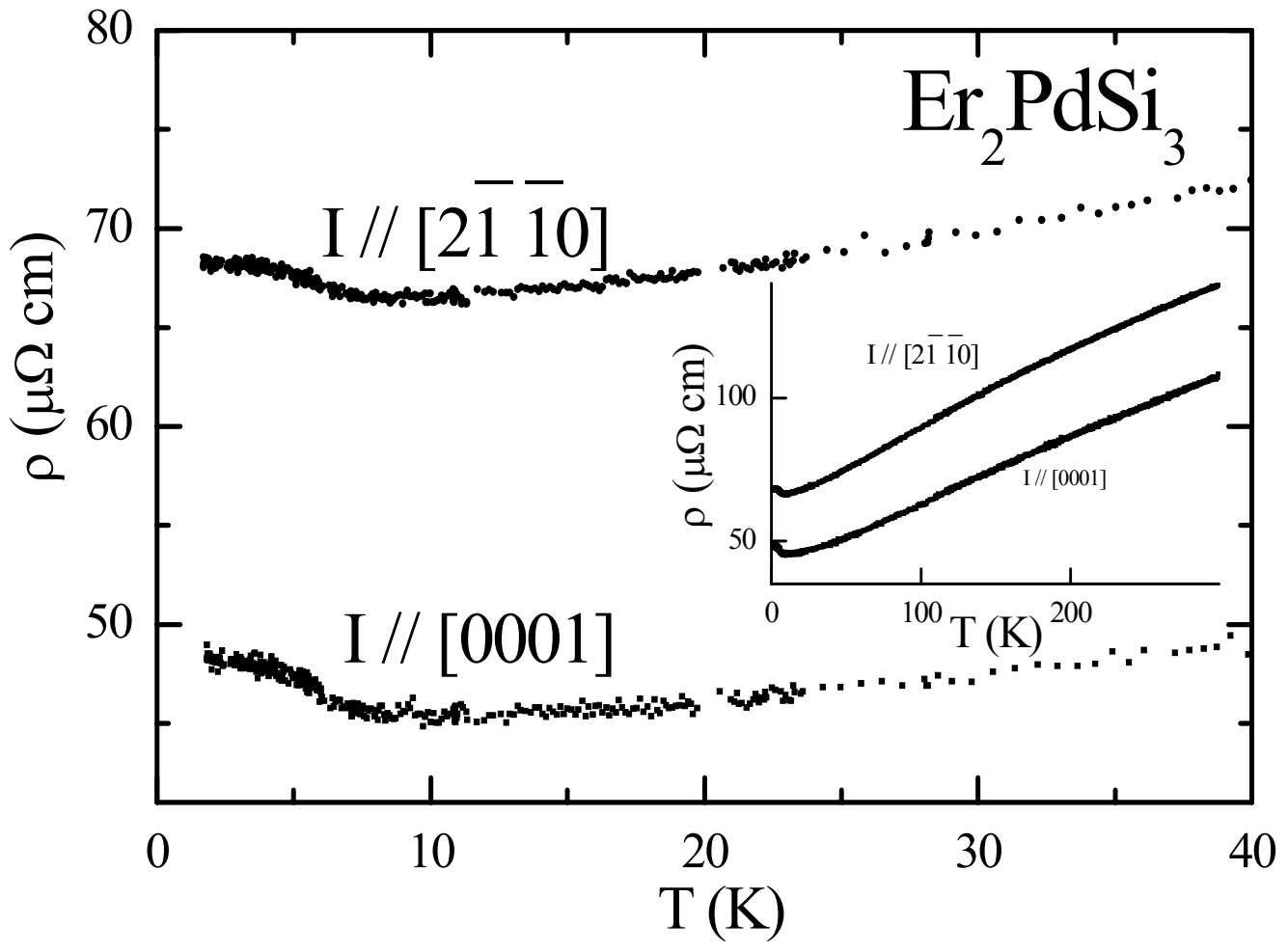

Figure 6. Electrical resistivity as a function of temperature for $Er_2PdSi_3$ for two different directions of excitation current (I) below 40 K. The data in the range extended to 300 K are shown in the inset.